\documentclass[onecolumn]{article}

\usepackage{graphicx}
\usepackage{amsmath}
\usepackage{epsf}
\usepackage{citesort}

\input{tcilatex}
\newcommand{\mynewcaption}[2]{\protect\caption{#2\label{#1}}}

\begin{document}

\author{A. van de Walle and M. Asta}
\title{First-Principles Investigation of Perfect and Diffuse Anti-Phase Boundaries
in HCP-Based Ti-Al Alloys}
\maketitle

\begin{abstract}
First-principles thermodynamic models based on the cluster expansion 
formalism, monte-carlo simulations and quantum-mechanical total
energy calculations are employed to compute short-range-order
parameters and diffuse-antiphase-boundary energies in hcp-based 
$\alpha$-Ti-Al alloys.  Our calculations unambiguously reveal a 
substantial amount of SRO is present in $\alpha$-Ti-6 Al and that, 
at typical processing temperatures concentrations, the DAPB energies 
associated with a single dislocation slip can reach 25 mJ/m$^{2}$. We 
find very little anisotropy between the energies of DAPBs lying in the 
basal and prism planes. Perfect antiphase boundaries in DO$_{19}$ ordered 
Ti$_3$Al are also investigated and their interfacial energies, interfacial 
stresses and local displacements are calculated from first principles 
through direct supercell calculations.  Our results are discussed in 
light of mechanical property measurements and deformation microstructure 
strudies in $\alpha$ Ti-Al alloys.
\end{abstract}

\section{Introduction}

The presence of chemical short-range order (SRO) in $\alpha $-Ti-Al was
proposed over thirty years ago by Blackburn and Williams \cite{Blackburn} to
explain planar slip exhibited by alloys with concentrations greater than
roughly 4 wt. \% Al (7 at. \% Al) \cite{Lim,Neeraj1}. Since that time SRO
has been discussed as an important factor contributing to the anomalous
solid-solution strengthening of $\alpha $-Ti-Al \cite
{Ogden,Blackburn,Rosenberg,Conrad}, as well as the low strain hardening
exponents exhibited by these alloys \cite{Ogden,Sakai,Neeraj1}. Only very
recently, however, has direct evidence for SRO been provided from diffuse
neutron scattering measurements \cite{Neeraj1,Neeraj2}. These measurements
confirm the presence of appreciable SRO in Ti-6 wt. \%Al alloys, of a type
consistent with the ordered D0$_{19}$ structure of $\alpha _{2}$ Ti$_{3}$Al.

As first proposed by Fisher \cite{Fisher}, SRO in solid solutions can
provide a significant source of strengthening due to the energy cost
associated with the creation of ``diffuse'' antiphase boundaries (DAPB)
arising from shearing of the lattice by dislocation slip. The DAPB is the
analogue for structures with SRO of the more familiar antiphase-boundary
(APB) in a long-range-ordered crystal structure. In solid solutions with
chemical SRO, the energy cost of creating a DAPB ($\gamma _{SRO}$, defined
as the excess energy per unit area of a DAPB) during slip gives rise to a
friction stress on moving dislocations: $\tau _{SRO}b=\gamma _{SRO}$. A plot
of the DAPB energy versus the number of shearing events, calculated from
first-principles for $\alpha $-Ti-Al alloys using the approach described
below, is shown in Fig. \ref{figgpt}. This plot is qualitatively similar to
theoretical results obtained for fcc-based alloy systems previously \cite
{Mohri1,Mohri2,Schwander,Plessing}. The energy of the DAPB after
the first shear (to be referred to as $\gamma _{1}$) is seen to be larger
than the values ($\gamma _{2}$, $\gamma _{3}$, etc.) of $\gamma _{SRO}$
after multiple shearing events. As pointed out originally by Cohen and Fine 
\cite{Cohen}, the second slip event in a solid-solution with SRO can
partially restore the state of order, giving rise to a decrease in $\gamma
_{SRO}$ (c.f. Fig. \ref{figgpt}) and a driving force for the pairing of
leading dislocations in pile-ups. Such pairing has been observed recently by
transmission electronc microscopy (TEM) for dislocation pile-ups in 
$\alpha $-Ti-Al \cite{Neeraj3}. Fig. \ref
{figgpt} also illustrates that after a small number of slip events the state
of order is largely destroyed and subsequent dislocations experience
negligible friction stress due to SRO. By this ``glide-plane-softening''
mechanism, SRO can promote planar slip in solid solutions, as proposed for $%
\alpha $-Ti-Al alloys by Blackburn and Williams \cite{Blackburn}.

Several TEM-based experimental studies have been undertaken to estimate the
magnitude of $\tau _{SRO}$ from observed dislocation structures \cite
{Prinz,Neuhauser,Clement,Olfe,Jouiad98,Jouiad99,Plessing,Neeraj1} in
short-range-ordered alloys. To date all of this work has been devoted to fcc
alloys, with the exception of the very recent experimental study by Neeraj 
\emph{et al.} \cite{Neeraj1} for hcp-based $\alpha $-Ti-Al. In this study,
magnitudes of $\gamma _{2}$ for DAPBs on prism planes were estimated from
the critical stress required to form screw dislocation dipoles from
elongated ``hair-pin'' loops, while the difference $\gamma _{1}-\gamma _{2}$
was estimated from measured spacings between paired leading edge
dislocations in pile-ups. This work represents the first direct measurement
of the friction stress due to SRO in $\alpha $-Ti-Al and is an important
contribution to the detailed understanding of deformation properties in
these alloys. The purpose of the research reported in this manuscript is
to provide an independent theoretical estimate of DAPB energies in this
system.

A number of theoretical studies have been undertaken to calculate the
magnitude of DAPB energies for fcc and bcc alloys \cite
{Flinn,Cohen,Cohen62,ROWilliams,Buchner,Inden,Mohri1,Mohri2,Schwander,Plessing,Pitsch74,Pitsch75}%
. These studies have employed related methods which require (i) a
description of the equilibrium state of short-range order in the alloy based
upon the SRO parameters, as well as (ii) values of the interatomic
interaction parameters which parametrize the energy associated with changes
in atomic configurations. From the SRO parameters the number of favorable
and unfavorable bonds created by a slip process can be derived and converted
to an excess energy per unit area from the values of the interaction
energies. To date the most comprehensive of such theoretical approaches was
developed by Schwander \emph{et al.} \cite{Schwander} who formulated general
expressions for $\gamma _{SRO}$ based upon an arbitrary number of SRO
parameters. For Ni-Cr alloys, Schwander \emph{et al.} made use of measured
SRO parameters to derive the magnitudes of pair interaction energies through
the inverse-Monte-Carlo method \cite{Gerold}. This approach was employed
more recently by Plessing \emph{et al.} \cite{Plessing} for Cu-Al and Cu-Mn
alloys, where good agreement was obtained between calculated values and
estimates of $\gamma _{SRO}$ derived from friction stresses deduced from
pile-up geometries in post-deformed samples. For $\alpha $-Ti-Al alloys,
measured values of the short-range-order parameters are currently
unavailable and the theoretical approach of Schwander \emph{et al.} cannot
be applied to the calculations DAPB energies in this system at present. An
alternative approach relies upon the use of quantum-mechanical calculations
to derive the values of the interatomic interaction parameters from
first-principles \cite{fontaine:clusapp,zunger:NATO}. These parameters can
be employed in statistical-mechanical calculations to compute values for
short-range-order parameters, allowing estimates of $\gamma _{SRO}$ to be
made without the need for experimental input. Such a first-principles
approach has been applied to the study of SRO hardening in Cu-Au alloys by
Mohri \emph{et al.} \cite{Mohri2}. In the present study first-principles
calculations of the composition and temperature dependencies of $\gamma
_{SRO}$ are undertaken for $\alpha $-Ti-Al alloys. To the best of our
knowledge, this work represents the first time a calculation of DAPB
energies has been undertaken for an hcp-based alloy system.

In the next section we begin by discussing first-principles results for the
structure, energetics and stress of APBs in ordered $\alpha _{2}$ Ti$_{3}$%
Al. The APB energy results serve as a basis for testing the predictive power
of the calculated model interatomic interaction parameters for hcp-based
Ti-Al alloys. The next section is devoted to a detailed discussion of our
study of DAPB energies. Our calculated values of $\gamma _{SRO}$ for DAPBs
on prism planes are found to be consistent with the recent experimental
estimates of Neeraj \emph{et al.} mentioned above.

\section{Anti-Phase Boundaries}

\subsection{Methodology}

The APB energies were calculated using the local density approximation (LDA)
to density functional theory (DFT), as implemented in the VASP package \cite
{kresse:vasp,kresse:vasp1,kresse:vasp2}, which relies on ultrasoft 
pseudopotentials \cite{Vanderbilt:soft_pseudo} and a plane wave basis. The plane wave
energy cutoff was set to 270 eV while the number of $k$-points was chosen to
be at least 3500/(atom$^{-1}$) for all structures.\footnote{%
e.g., (3500/$n$) $k$-points would be selected for a structure with $n$ atom
per unit cell.} A supercell geometry was used, where a periodic array of APBs
is set up. It was verified that the APBs are sufficiently spaced from one
another to avoid interactions by computing the APB energy for different
supercell sizes. Separations of $5$ and $16$ monolayers were considered.
Atomic positions were fully relaxed and the dimensions of the supercell 
perpendicular to the plane of the APB were allowed to relax to give zero
normal stress. The lattice parameters along the
plane of the interface were frozen to the equilibrium lattice parameter of
the DO$_{19}$ structure.

The error bars we provide were derived by studying the convergence of our
results with respect to plane wave energy cutoff, number of $k$-points,
supercell size and the tolerance for residual forces and stresses for the
numerical determination of the equilibrium geometry.

\subsection{Results and Discussion}

We consider the APB lying in the basal $\left[ 0001\right] $ and prism $%
\left[ 1\overline{1}00\right] $ planes, as these are the main active slip
systems in Ti-Al alloys. The calculated APB energies are reported in Table 
\ref{tabapb}. While there is only one type of basal APB, there are two types of
prism APB, differing by the location of the APB plane, as illustrated in
Fig. \ref{figp12}.

As shown in Table \ref{tabapbcomp}, the calculated APB energies\ are
comparable to the result of earlier all-electron first-principles
calculations \cite{zou:ti3al}, indicating that the pseudopotential
approximation used provides reliable results. Our energies are
systematically smaller, however, perhaps due to a more thorough treatment of
the relaxations made possible by the high computational efficiency of
pseudopotential-based methods.

Table \ref{tabapbcomp} also shows that the APB energies calculated from
first principles differ markedly from experimental estimates \cite{legros:apb},
which were obtained from measurements of the spacing between
dissociated dislocations, inferring the interface energy
through mechanical equilibrium considerations involving the trade-off
between elastic and interfacial energy. As discussed in the next section,
we estimate that the accuracy of
first-principles calculations may be responsible for an overestimation 
of at most 40\% of the APB\ energies.  This error is significantly 
smaller that the observed disagreement with experimental estimates. 
 The level of discrepancy between theory and experiment displayed in 
Table \ref{tabapbcomp} has been observed for related intermetallic 
systems \cite{AstaQuong,Marcel}.  While the source of this discrepancy 
remains unclear, we note that the equilibrium spacings implied by 
our APB energies (using the same elastic constants as in \cite{legros:apb}) 
are 3.0 nm, 7.0 nm, and 1.6 nm,\ respectively, for the basal, prism I, and 
prism II APB. These values are all very close to the \emph{minimum} 
dislocation spacings measured by \cite{legros:apb}.

The calculated APB energies are consistent with an intuitive bond-breaking
picture. The prism I, basal and prism II APB respectively create 0,1 and 2
Al-Al bonds, implying that the basal APB should have an energy than is
approximately equal to the average of the prism I and prism II APB ($293$
mJ/m$^{2}$).\footnote{%
The atomic density in the plane of the APB is slightly different in basal
and prism APB so that even in an ideal bond-break picture, the basal APB\
energy would differ slightly from the average energy of the prism APBs.} Our
calculations also provide interface stress and displacements (see Table \ref
{tabapb}). An analysis of their sign and magnitude reveals interesting
features of the relaxation taking place in the vicinity of an APB in Ti$_{3}$%
Al which may help in interpreting the stress and displacement fields that
can be quantified with TEM. For the two APBs that result in a change of the
nearest neighbor environment (basal and prism II), the interface contracts
perpendicular to the plane of the APB ($d_{1}<0$) and a tendency to expand
along the plane of the interface ($\sigma _{22}>0$ and $\sigma _{33}>0$).
This behavior is reversed for the prism I APB, where no change in the
nearest-neighbor environment occurs. The fact that, in all cases, the stress
along the interface and the displacement perpendicular to it have opposite
signs can be understood from a volume conservation argument. However, we
were unable to rationalize the relative magnitudes of the stresses and
displacements on the basis of intuitive atomic size considerations. This is
not surprising given that such a picture is unable to describe the
differences in bond length between hcp Ti and (metastable) hcp Al. According
to our first-principles calculations, in these two pure elements, the Ti-Ti
bonds are longer than Al-Al bonds lying along the (0001) plane while the
opposite is true for other nearest neighbor bonds.

\section{Diffuse Anti-Phase Boundaries}

\subsection{Methodology}

The calculation of diffuse anti-phase boundary (DAPB) energies requires the
construction of a thermodynamic model of the alloy. The cluster expansion
formalism \cite{sanchez:cexp,fontaine:clusapp,zunger:NATO} was used to
parametrize the configurational dependence of the energy of the alloy
system: 
\begin{equation}
E\left( \sigma \right) =\sum_{i}J_{i}\sigma _{i}+\sum_{\left\{ i,j\right\}
}J_{ij}\sigma _{i}\sigma _{j}+\sum_{\left\{ i,j,k\right\} }J_{ijk}\sigma
_{i}\sigma _{j}\sigma _{k}+\ldots  \label{cemain}
\end{equation}
where $\sigma _{i}$ denotes the occupation of site $i$ ($\sigma _{i}=1$ for
Al, $\sigma _{i}=-1$ for Ti) while the $J_{i\ldots }$ are effective
interaction parameters to be determined from first principles and where the
summations are taken over every singlet, pair, triplet,... of lattice sites.
The so-called effective cluster interactions (ECI) $J_{i\ldots }$ embody all
the contributions to the energetics of the alloy, including relaxations of
the atoms away from their ideal lattice sites. These ECI were fitted to the
energies of a set of $55$ periodic structures calculated from
first-principles using the same techniques and parameters as for the APB
calculations. This procedure was carried out using the MAPS software package 
\cite{avdw:maps} which automates most of the tasks required to apply the
cluster expansion formalism.

The ECI provide all the information we need to compute the DAPB energies.
First, the ECI can be directly used to perform Monte Carlo simulations \cite
{binder:mc}\ that allow a first-principles computation of the phase diagram 
of the alloy system and degree of short-range order present in the alloy 
at a given temperature and
composition (\cite{Wolverton}, and references therein). To this effect, we
set up a Monte Carlo simulation cell consisting of a $12\times 12\times 6$
supercell of the hcp unit cell. The initial configuration was chosen to be
random and $15000$ equilibration Monte Carlo passes were performed before
the short-range order was computed during an additional $2500$ passes. The
size of simulation cell and the number of simulation steps are such that the
statistical noise in the computed short-range order parameters is orders of
magnitude smaller than the error arising from the precision of the ECI
themselves.

The DAPB energies $E_{DAPB}$ per unit area can then be directly expressed in
terms of the ECI and the short-range order parameters obtained from the
Monte Carlo simulation: 
\begin{equation}
E_{DAPB}=\frac{1}{A}\sum_{\left\{ i,j\right\} }\left( \left\langle \sigma
_{i}\sigma _{j}\right\rangle -\left\langle \sigma _{S\left( i\right) }\sigma
_{S\left( j\right) }\right\rangle \right) J_{ij}+\ldots   \label{linapb}
\end{equation}
where $S$ is a function that maps a given site $i$ onto its new position,
after the anti-phase boundary has been created and where $\left\langle
\sigma _{i}\sigma _{j}\right\rangle $ denotes the thermodynamic and spatial
average of the correlation between the occupation of sites $i$ and $j$. The
sum is taken over all pairs (triplets, etc.) that are not equivalent under a
translation lying in the DAPB plane which leaves the lattice invariant. $A$
is the area of a repeat unit of the plane of interest. This formulation,
which is equivalent to the one introduced by Schwander \cite{Schwander}, has
the advantage that it makes the most efficient use of all the information
provided by the Monte Carlo simulations. It allows us to obtain DAPB
energies \emph{without} having to average over various DAPB energy obtained
by running a dislocation through various snapshots of the Monte Carlo cell
during the simulation. This averaging is implicitly taken into account
through the averaging of the correlations $\left\langle \sigma _{i}\sigma
_{j}\right\rangle $ over the Monte Carlo steps and over all symmetrically
equivalent clusters. As the cluster expansion does not directly provide
information regarding the area $A$, we assume the disordered phase to be an
hcp phase with the idea $c/a$ ratio and with an average atomic volume equal
to the one of the DO$_{19}$ structure. For a system with a small size
mismatch ($3$\%, in our case), the error in $A$ introduced by this
approximation will amount to at most $6$\%.

Since we have access to the energy of a finite number of structures, we can
only determine a finite number of ECI and the expansion defined by Equation (%
\ref{cemain}) needs to be truncated. The choice of the optimal number of
terms to keep was made through the following statistical analysis. As our
goal is ultimately to predict DAPB energies, we seek to construct a cluster
expansion that exhibits a good predictive power for alloy configurations
similar to perfect APB. This exercise does not merely consists in trying
many different ranges of interaction until one happens to predict APB
energies that agree with the calculated ones. The agreement could be
fortuitous and the predictive power of the cluster expansion for
configurations that are slightly different from a perfect APB, such as a
DAPB, could be poor. Instead, we calculate the statistical variance of the
predicted APB energies for various choices of range of interaction, and
look for the choice that minimizes that variance. The statistical variance
is computed as follows. Let $J$ denote the vector of all the selected ECI
and let $V$ be their estimated covariance matrix.\footnote{%
Let $Y$ be an $n$-dimenstional vector and $X$ be a $n\times k$ matrix. From
standard statistical analysis, the covariance matrix of a parameter vector $%
b $ obtained from a least square fit $\min_{b}\left\{ \left( Y-Xb\right)
^{T}\left( Y-Xb\right) \right\} $ can be estimated by $\left( X^{T}X\right)
^{-1}\sigma ^{2}$ where $\sigma ^{2}$ is the mean square error of the fit.}
The energy of a given APB, $E_{APB}$ is related to the vector $J$ of all ECI
by Equation (\ref{linapb}), where the correlations $\left\langle \sigma
_{i}\sigma _{j}\right\rangle $ are calculated for a perfect DO$_{19}$
structure. This linear relationship can be written as, $E_{APB}=C^{T}J$,
where $C$ is some vector. As a result, the variance matrix of the predicted
APB energy is then simply given by: $C^{T}VC$.

\subsection{Assessment of the Cluster Expansion}

The ECI of the cluster expansion that were found to minimize the sum of the
statistical variances of the predicted APB energies are listed in Table \ref
{tabeci}. The ECI we calculated are very close to the ones that were
determined in \cite{shimono:hcpeci} by a fit to the experimental Ti-Ti$_{3}$%
Al phase boundary. The resulting cluster expansion predicts APB energies
that agree remarkably well with the calculated APB (see Table \ref{predapb}%
), given that the perfect APB energies were not used in the fit of the
cluster expansion. The difference between the calculated and predicted
energies is of the same order of magnitude as the error bars, confirming
that our variance calculations indeed provide a very good measure of the
true predictive power of the cluster expansion. A similar variance analysis
will be used in the next section to estimate the errors on our calculated
DAPB energies.

Although the selected range of interaction is by construction the one that
will best predict the APB energies, it is still possible that longer range
interactions may have an important effect on the calculated thermodynamic
properties of the alloy. To investigate this concern, longer range
interactions were gradually added. Adding up to $8$ additional pair
interactions and $6$ triplet interactions changed the temperature of the DO$%
_{19}\rightarrow $hcp transition by less than 200K and we are thus confident
that our short-range cluster expansion is sufficiently accurate for our
purposes.

Table \ref{ce_param} gives a few characteristics of our cluster expansion.
The precision of the cluster expansion is quite good, as can be seen by the
value of both the mean square error and the cross-validation score.%
\footnote{%
The cross-validation score is defined to be $\left( \frac{1}{n}%
\sum_{s=1}^{n}\left( E_{s}-\hat{E}_{s}\right) ^{2}\right) ^{1/2},$ where $%
E_{s}$ is the calculated energy of structure $s$ and $\hat{E}_{s}$ is the
value of the same energy predicted using the cluster expansion fitted to the
remaining $n-1$ structures.}

The temperature $T_{c}$ of the DO$_{19}\rightarrow $hcp transition was
calculated to be $2000$K, which is about 40\% higher than the known
experimental transition temperature of $1420$K. Such an overestimation of
transition temperatures is quite common in first-principles calculations
based on the local-density approximation and short-ranged cluster expansions
(a list of systems where such first-principles phase diagram calculations
have been undertaken can be found in \cite{fontaine:clusapp}). To correct
for this overestimation of $T_{c}$, we normalize all our temperatures by $%
T_{c}$. Experimental measurements should then be compared with the
calculations performed at the same normalized temperature. In Fig. \ref
{figpb}, the Ti-rich portion of the calculated phase diagram is seen to
agree very well with the experimentally measured one, when both are
normalized to have the same $T_{c}$.

The error in the calculated transition temperature provides us with an
estimate of the error introduced by (i) the use of the Local Density
Approximation in the first-principles calculations (ii) neglecting
non-configurational sources of entropy, such as lattice vibrations and
electronic entropy, and (iii) inaccuracies in the description of the
configurational free energy arising from the truncation of the cluster
expansion to a finite number of terms (see above). We thus expect our
perfect and diffuse APB energies to be biased upward by at most 40\% (this
correction has neither been applied to our result nor been included in the error
bars, in order to allow comparison with other first-principles results).

\subsection{Results and Discussion}

Our first main result is that Ti-rich Ti-Al alloys do indeed exhibit a
substantial amount of short-range order, as shown in Fig. \ref{figsro},
which plots the Warren-Cowley short-range order parameters \cite{Warren} as
a function of distance. This short-range order is of the DO$_{19}$ type. Figures 
\ref{figgpt} and \ref{figgpx} show the calculated values of the DAPB
energies $\gamma _{k}$ for the prism DAPB for the temperatures and
concentrations marked by a cross in Fig. \ref{figpb}. Energies of the basal
DAPB are not shown, as they are nearly identical to those for the prism
DAPB, as illustrated in Fig. \ref{figgpb} for one specific temperature and
concentration. In these plots, the abscissa $k$ represents the magnitude of
the shift (in multiples of lattice vectors) between the two half crystals on
each side of the DAPB, or, alternatively, the number of dislocations needed
to create the DAPB. Error bars are not shown for clarity, but amount to
about 15\% of the calculated DAPB energies resulting from the
statistical noise in the fitted ECI.

Note that the two types of prism APB can no longer be distinguished in a
short-range ordered alloy because the DAPB plane cuts through various
short-range ordered domains which have a random relative phase, resulting in
a single average interfacial energy. The plots exhibit the characteristic
oscillations that are expected from the observation that, in a perfectly
ordered DO$_{19}$ structure, every other dislocation would restore a perfect
state of order. In a short-range ordered alloy, the ability of subsequent
dislocations to restore the original state of order decreases with the number
of dislocations as a result of the short-range nature of the state of order.

Our estimates of the DAPB energies are comparable to those found
experimentally \cite{Neeraj1}. For $\alpha $ Ti-6 wt. \% Al alloys annealed
at temperatures between 600 and 350$%
{{}^\circ}%
$C, experimental estimates of $\gamma _{2}$ ranged between 25 and 48 mJ/m$%
^{2}$, while a value of 6 mJ/m$^{2}$ was quoted for the difference $\gamma
_{1}-\gamma _{2}$. Our results indicate that for an alloy of the same
composition (10 at. \% Al) and for the same range of relative temperature ($%
T/T_{c}$ ranging from $0.45$ to $0.6$), $\gamma _{2}=15$ mJ/m$^{2}$ while $%
\gamma _{1}-\gamma _{2}$ ranges from $7$ mJ/m$^{2}$ to $10$ mJ/m$^{2}$. 
These values of $\gamma _{2}$ are within a factor of 2 relative
to experimental results, while our estimates of the difference $\gamma
_{1}-\gamma _{2}$ are comparable.

An important finding is that for a substantial portion of the region of
solid solubility of Al in Ti, the DAPB energies $\gamma _{1}$ are larger
than $2-4$ mJ/m$^{2}$, values which have been quoted \cite{Roelofs} as the
thresholds above which transitions from homogenous to planar slip occur in
fcc-based Cu-Zn, Cu-Al, and Cu-Mn alloys. This result supports the
proposition that the presence of SRO contributes to the predominantly
planar slip in Ti-Al \cite{Blackburn}.

The fact that prism and basal DAPB have an almost identical energy is again
consistent with a simple bond-breaking picture.\footnote{%
This is not an artifact of the use of a short-range cluster expansion. The
inclusion of second nearest-neighbor interactions allows the basal and prism
DAPB energies to differ.} Despite the small anisotropy in the average DAPB
energies, the behavior of each type of DAPB under stress may differ. The
local DAPB energy density may very well fluctuate as we move along the
interface. For instance, a prism DAPB may contain portions that are locally
analogous to a low energy prism I APB and portions that are locally similar
to a high energy prism II APB, while such fluctuations should be much
smaller in basal DAPB where only one type of local order exists. The average
DAPB energy we compute is the quantity of interest in order to determine,
for instance, the total work required to deform the material by a given
amount. However, for properties such as the critical resolved shear stress
(CRSS), the average may not provide enough information. A more detailed
investigation of the spatial fluctuations in DAPB energy density would be
needed, a topic we leave for future research.

Our calculated $\gamma _{k}$ are of the same order of magnitude as earlier
theoretical studies which focused on fcc alloys \cite
{Mohri1,Mohri2,Schwander,Plessing}. One notable qualitative distinction is
the fact that, in hcp Ti-Al alloys, the second dislocation restores the
original state of short-range order to a greater extent than in these fcc
alloys, as can be seen by the larger drop between $\gamma _{1}$ and $\gamma
_{2}$. We attribute this difference to the fact that, in all the fcc alloys
studied, the thermodynamically stable ordered phase in the vicinity of the
short-range ordered region is a DO$_{22}$ structure. Due to the tetragonal
symmetry of this phase, different short-range ordered domains can have
distinct orientations. Since the DAPB cuts through these different domains
and since the DO$_{22}$ has different periodicity in different directions,
it is impossible for the second dislocation to restore the local order in
all domains simultaneously. In contrast, in the case of Ti-Al, the hcp-based
DO$_{19}$ structure has the same periodicity along any $\left[ 1%
\overline{1}00\right] $ direction and it follows that the second dislocation can
partially restore the original local order in all domains simultaneously.

\section{Conclusion}

Our first main result is the confirmation, on the basis of first-principles
calculations, that DO$_{19}$-type short-range is present in the hcp $%
\alpha $-Ti phase of the Ti-Al alloy.  We have also calculated the energy 
of diffuse anti-phase boundaries (DAPB). We find that for an alloy at a 
composition of 10 at. \% of Al and a temperature between 600 and 300$%
{{}^\circ}%
$C, a single dislocation creates a DAPB with an interfacial energy ranging
from $22$ to $25$ mJ/m$^{2}$ while a second dislocation creates a DAPB
having an interfacial energy of $15$ mJ/m$^{2}$. These values are
significantly higher than $4$ mJ/m$^{2}$, a value estimated to be sufficient to
induce a transition from homogeneous to planar slip in fcc-based alloys \cite
{Roelofs}.

The computer programs used to complete this work can be obtained by
contacting the authors. These include an easy-to-use Monte Carlo code that
provides, among other quantities, short-range order parameters and automatic
phase boundary plotting capabilities. A companion code implements the
Schwander \textit{et al. }\cite{Schwander} formalism to its full generality,
allowing the calculation of DAPB of any orientation, for any lattice type,
and for any number of short-range order parameter.

\section*{Acknowledgements}

We are grateful for numerous helpful discussions with Dr. T. Neeraj and
Prof. M. Mills from The Ohio State University who suggested this problem,
and who provided us with a copy of the dissertation cited in Ref. \cite
{Neeraj1}. This research was supported by the National Science Foundation
under program DMR-0080766.

\newpage


\begin{thebibliography}{10}

\bibitem{Blackburn}
M.~J. Blackburn and J.~C. Williams, Trans. ASM {\bf 62}, pp. 398--399  (1969).

\bibitem{Lim}
J.~Y. Lim, J. McMahon, D.~P. Pope, and J.~C. Williams, Metall. Trans. {\bf 7A},
  pp. 139--144  (1976).

\bibitem{Neeraj1}
T. Neeraj, Ph.D. thesis, The Ohio State University, 2000.

\bibitem{Ogden}
H.~R. Ogden, D.~J. Maykuth, W.~L. Finley, and R.~I. Jaffee, Trans. AIME {\bf
  197}, pp. 267--272  (1953).

\bibitem{Rosenberg}
H.~W. Rosenberg and W.~D. Nix, Metall. Trans. {\bf 4}, pp. 1333--1342  (1972).

\bibitem{Conrad}
H. Conrad, Scripta Metall. {\bf 7}, pp. 509--512  (1973).

\bibitem{Sakai}
T. Sakai and M.~E. Fine, Scripta Metall. {\bf 8}, pp. 541--544  (1974).

\bibitem{Neeraj2}
T. Neeraj, J.~L. Robertson, and M.~J. Mills (unpublished).

\bibitem{Fisher}
J.~C. Fisher, Acta Metall. {\bf 2}, pp. 9--10  (1954).

\bibitem{Mohri1}
T. Mohri, D. de~Fontaine, and J.~M. Sanchez, Metall. Trans. {\bf 17A}, pp. 189--194
  (1986).

\bibitem{Mohri2}
T. Mohri, T. Tsutsumi, O. Sasaki, and K. Watanabe, Metall. Trans. {\bf 21A},
 pp. 3165--3169  (1990).

\bibitem{Schwander}
P. Schwander, B. Sch{\"{o}}nfeld, and G. Kostorz, Phys. Stat. Sol. (b) {\bf
  172}, pp. 73--85  (1992).

\bibitem{Plessing}
J. Plessing, C. Achmus, H. Neuhauser, B. Sch\"{o}nfeld, and G. Kostorz,
Z. Metallkd. {\bf 88}, pp. 630--635  (1997).

\bibitem{Cohen}
J.~B. Cohen and M.~E. Fine, Acta Metall. {\bf 11}, pp. 1106--1115  (1963).

\bibitem{Neeraj3}
T. Neeraj, D.-H. Hou, G.~S. Daehn, and M.~J. Mills, Acta Mater. {\bf 48}, pp. 1225--1238
   (2000).

\bibitem{Prinz}
F. Prinz, H.~P. Karnthaler, and H.~O.~K. Kirchner, Acta Metall. {\bf 29}, pp. 1029--1036
   (1981).

\bibitem{Neuhauser}
H. Neuha{\"{u}}ser, O.~B. Arkan, and H.~H. Potthoff, Mater. Sci. Eng. {\bf 81},
 pp.  201--209  (1986).

\bibitem{Clement}
N. Clement, D. Caillard, and J.~L. Martin, Acta. Metall. {\bf 32}, pp. 961--975
  (1984).

\bibitem{Olfe}
J. Olfe and H. Neuh{\"{a}}user, phys. stat. sol. (a) {\bf 109}, pp. 149--160  (1988).

\bibitem{Jouiad98}
M. Jouiad, N. Clement, and A. Coujou, Phil. Mag. A {\bf 77}, pp. 689--699  (1998).

\bibitem{Jouiad99}
M. Jouiad, F. Pettinari, N. Clement, and A. Coujou, Phil. Mag. A {\bf 79},
 pp. 2591--2602  (1999).

\bibitem{Flinn}
P.~A. Flinn, Acta Metall. {\bf 6}, pp. 631--635  (1958).

\bibitem{Cohen62}
J.~B. Cohen and M.~E. Fine, J. Phys. et Radium {\bf 23}, pp. 749--762  (1962).

\bibitem{ROWilliams}
R.~O. Williams, Acta Metall. {\bf 18}, pp. 457--466  (1970).

\bibitem{Buchner}
A.~R. B{\"{u}}chner and W. Pitsch, Z. Metallkd. {\bf 76}, pp. 651--656  (1985).

\bibitem{Inden}
G. Inden, S. Bruns, and H. Ackermann, Phil. Mag. A {\bf 53}, pp. 87--100  (1986).

\bibitem{Pitsch74}
W. Pitsch, Scripta Metall. {\bf 8}, pp. 813--819  (1974).

\bibitem{Pitsch75}
W. Pitsch, Scripta Metall. {\bf 9}, pp. 1059--1062  (1975).

\bibitem{Gerold}
V. Gerold and J. Kern, Acta Metall. {\bf 35}, pp. 393--399  (1987).

\bibitem{fontaine:clusapp}
D. de~Fontaine, Solid State Phys. {\bf 47}, pp. 33--176  (1994).

\bibitem{zunger:NATO}
A. Zunger,  in {\em NATO ASI on Statics and Dynamics of Alloy Phase
  Transformation}, edited by P.~E. Turchi and A. Gonis (Plenum Press, New York,
  1994), Vol.~319, pp. 361--405.

\bibitem{kresse:vasp}
G. Kresse and J. Furthm{\"u}ller, Vienna Ab-initio Simulation Package (VASP).

\bibitem{kresse:vasp1}
G. Kresse and J. Furthm{\"u}ller, Phys. Rev. B {\bf 54}, pp. 11169--11186  (1996).

\bibitem{kresse:vasp2}
G. Kresse and J. Furthm{\"u}ller, Comp. Mat. Sci. {\bf 6}, pp. 15--50  (1996).

\bibitem{Vanderbilt:soft_pseudo}
D. Vanderbilt, Phys. Rev. B {\bf 41}, pp. 7892--7895  (1990).

\bibitem{zou:ti3al}
C.~L. Fu, J. Zou, and M.~H. Yoo, Scripta Metall. Mater. {\bf 33},
 pp. 885--891  (1995).

\bibitem{legros:apb}
M. Legros, A. Couret, and D. Caillard, Phil. Mag. A {\bf 73}, pp. 61--80  (1996).

\bibitem{AstaQuong}
M. Asta and A.~A. Quong, Phil. Mag. Lett. {\bf 76}, pp. 331--339  (1997).

\bibitem{Marcel}
M.~H.~F. Sluiter, Y. Hashi, and Y. Kawazoe, Comput. Mater. Sci. {\bf 14}, pp. 283--290
  (1999).

\bibitem{sanchez:cexp}
J.~M. Sanchez, F. Ducastelle, and D. Gratias, Physica {\bf 128A}, pp. 334--350  (1984).

\bibitem{avdw:maps}
A. van~de Walle and G. Ceder, MIT Ab-initio Phase Stability (MAPS) code, available by
  contacting avdw{@}alum.mit.edu.

\bibitem{binder:mc}
K. Binder and D.~W. Heermann, {\em Monte Carlo Simulation in Statistical
  Physics} (Springer-Verlag, New York, 1988).

\bibitem{Wolverton}
C. Wolverton, V. Ozoli\c{n}\v{s}, and A. Zunger, J. Phys.: Condens. Matter {\bf
  12}, pp. 2749--2768  (2000).

\bibitem{shimono:hcpeci}
M. Shimono and H. Onodera, Phys. Rev. B {\bf 61}, pp. 14271--14274  (2000).

\bibitem{Warren}
B.~E. Warren, {\em X-Ray Diffraction} (Addison-Wesley, Reading, MA, 1969).

\bibitem{Roelofs}
H. Roelofs, B. Sch{\"{o}}nfeld, G. Kostorz, and W. Buhrer, phys. stat. sol. (b)
  {\bf 187}, pp. 31--42  (1995).

\bibitem{kattner:tial}
U.~R. Kattner, J.-C. Lin, and Y.~A. Chang, Metall. Trans. A {\bf 23A}, pp. 2081--2090
  (1992).

\end{thebibliography}

\clearpage

\section*{Tables}

\begin{table}[ht] \centering%
%
\begin{tabular}{|r|c|c|c|}
\hline
& basal & prism I & prism II \\ \hline
$\gamma $ (mJ/m$^{2}$) & \multicolumn{1}{|r|}{$249\pm 5$} & 
\multicolumn{1}{|r|}{$108\pm 2$} & \multicolumn{1}{|r|}{$477\pm 5$} \\ \hline
$\sigma _{22}$ (J/m$^{2}$) & \multicolumn{1}{|r|}{$1.4\pm 0.2$} & 
\multicolumn{1}{|r|}{-$1.5\pm 0.2$} & \multicolumn{1}{|r|}{$3.8\pm 0.2$} \\ 
\hline
$\sigma _{33}$ (J/m$^{2}$) & \multicolumn{1}{|r|}{$0.8\pm 0.2$} & 
\multicolumn{1}{|r|}{-$0.1\pm 0.2$} & \multicolumn{1}{|r|}{$0.3\pm 0.2$} \\ 
\hline
$d_{1}\text{ (\AA )}$ & \multicolumn{1}{|r|}{-$0.02\pm 0.01$} & 
\multicolumn{1}{|r|}{$0.16\pm 0.01$} & \multicolumn{1}{|r|}{-$0.10\pm 0.01$}
\\ \hline
$d_{3}\text{ (\AA )}$ & -$0.15\pm 0.03$ & $0$ & $0$ \\ \hline
\end{tabular}
\caption{Calculated Interface Energy $\gamma$, Interface Stress $\sigma$
and Displacement $d$ of the Basal and Prism APB.
Stresses and displacements are reported in cartesian coordinate system where the two first basis vectors are
respectively defined by the unit normal and the Burgers vector of the APB. Compressive stresses
are positive. The negative sign of $d_{3}$ indicates that  Al-Al bonds at the interface are shortened
by the displacements. All unreported stesses and displacements are zero by symmetry.
\label{tabapb}}%
\end{table}%
%

\begin{table}[ht] \centering%
%
\begin{tabular}{|r|c|c|c|}
\hline
& \multicolumn{3}{|c|}{$\gamma $ (mJ/m$^{2}$)} \\ \hline
& basal & prism I & prism II \\ \hline
present & \multicolumn{1}{|r|}{$249$} & \multicolumn{1}{|r|}{$108$} & 
\multicolumn{1}{|r|}{$477$} \\ \hline
first-principles \cite{zou:ti3al} & \multicolumn{1}{|r|}{$300$} & 
\multicolumn{1}{|r|}{$133$} & \multicolumn{1}{|r|}{$506$} \\ \hline
experiment \cite{legros:apb} & \multicolumn{1}{|r|}{$63$} & 
\multicolumn{1}{|r|}{$42$} & \multicolumn{1}{|r|}{$84$} \\ \hline
\end{tabular}
\caption{Comparison between calculated and Measured Interface Energy $\gamma$ of the Basal and Prism APB.
\label{tabapbcomp}}%
\end{table}%
%

\begin{table}[ht] \centering%
%
\begin{tabular}{|c|r|r|}
\hline
Pair & Interaction (eV) & Ref \cite{shimono:hcpeci} \\ \hline
$\left( 0,0,0\right) -\left( \frac{2}{3},\frac{1}{3},\frac{1}{2}\right) $ & $%
0.057\pm 0.005$ & $0.0588$ \\ \hline
$\left( 0,0,0\right) -\left( 1,0,0\right) $ & $0.047\pm 0.003$ & $0.0588$ \\ 
\hline
$\left( 0,0,0\right) -\left( \frac{2}{3},-\frac{2}{3},\frac{1}{2}\right) $ & 
$-0.014\pm 0.004$ & $-0.0235$ \\ \hline
\end{tabular}
\caption{ECI of the Cluster Expansion. Coordinates are expressed in the
conventional hexagonal coordinate system. Error bars are one standard
deviation wide.\label{tabeci}}%
\end{table}%
%

\begin{table}[ht] \centering%
%
\begin{tabular}{|l|l|l|l|}
\hline
$\gamma $ (mJ/m$^{2}$) & basal & prism I & prism II \\ \hline
supercell & $249\pm 5$ & $108\pm 2$ & $477\pm 5$ \\ \hline
clus. exp. & $234\pm 34$ & $71\pm 20$ & $427\pm 55$ \\ \hline
\end{tabular}
\caption{APB Energies Obtained from Supercell Calculations and Predicted
from a Cluster Expansion. Error bars on the predicted energies are one
standard deviation wide.\label{predapb}}%
\end{table}%
%

\begin{table}[ht]\centering%
%
\begin{tabular}{|l|l|}
\hline
Number of structures: & $55$ \\ \hline
Atom/unit cell: & $2-12$ \\ \hline
Mean square error: & $0.031$ eV/atom \\ \hline
Cross-validation score: & $0.034$ eV/atom \\ \hline
Predicted $T_{c}$ of DO$_{19}\rightarrow $hcp: & $2000$K$\pm 25$K \\ \hline
Experimental $T_{c}$ of DO$_{19}\rightarrow $hcp: & $1420$K \\ \hline
\end{tabular}
\caption{Characteristics of the Cluster Expansion.\label{ce_param}}%
\end{table}%
%

\clearpage

\section*{Figures}

\begin{figure}[ht]%
%
\centerline{\epsfbox{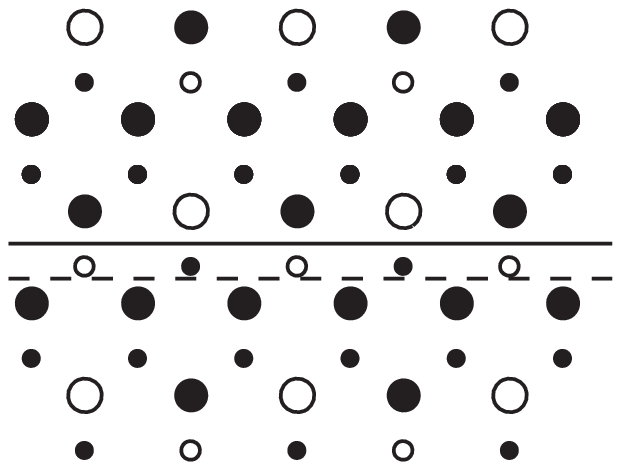}}%
\mynewcaption{figp12}{DO$_{19}$ structure and the location of prism I (dashed line) and prism II (solid line) planes.
The size of the circles distinguishes distinct (0001) planes of atom. Al atoms are shown in white, Ti atoms in black.}%
\end{figure}%
%

\begin{figure}[ht]%
%
\centerline{\epsfbox{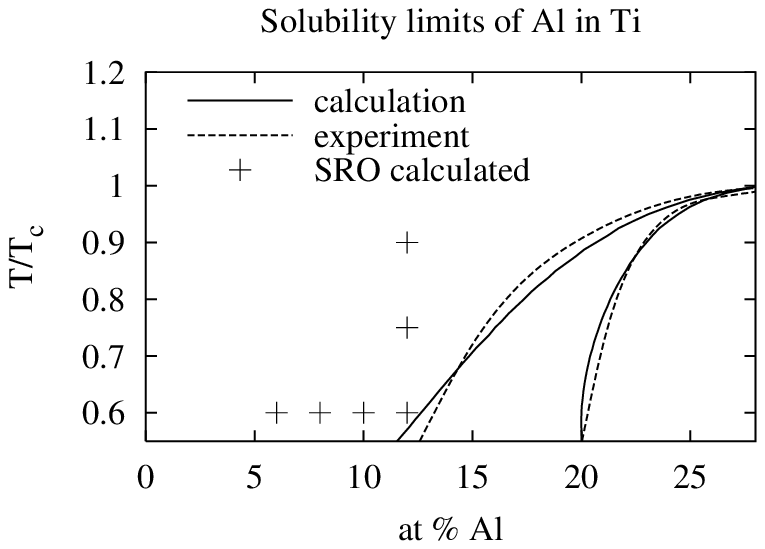}}%
\mynewcaption{figpb}{Calculated and Experimental Phase Boundaries (from \cite{kattner:tial}).
Also shown are the points where the DAPB energies were calculated.}%
\end{figure}%
%

\begin{figure}[ht]%
%
\centerline{\epsfbox{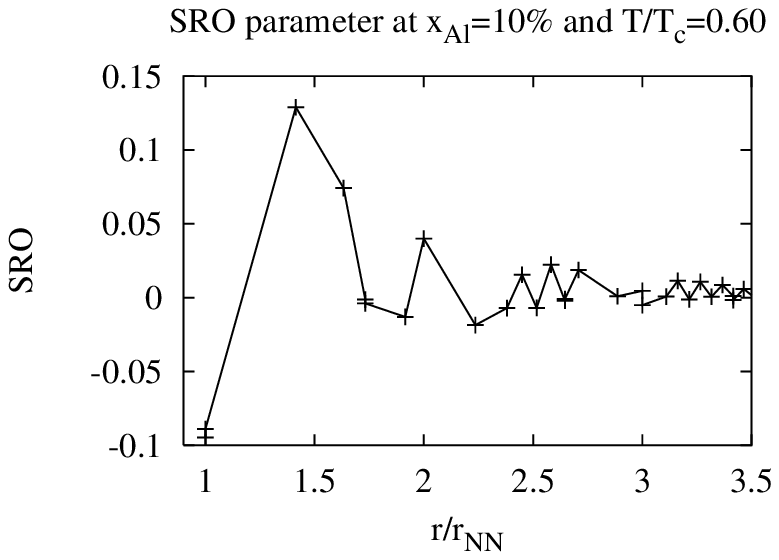}}%
\mynewcaption{figsro}{Calculated Warren-Cowley short-range order parameter as a function of distance
(normalized by nearest neighbor distance).}%
\end{figure}%
%

\begin{figure}[ht]%
%
\centerline{\epsfbox{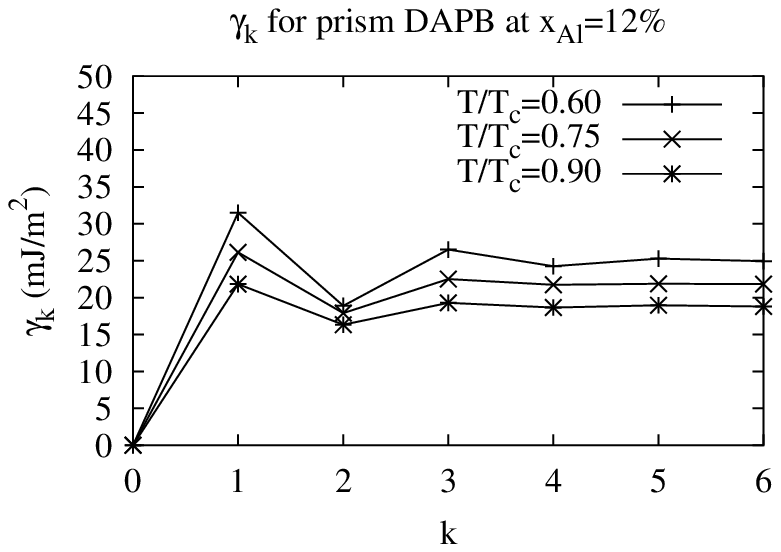}}%
\mynewcaption{figgpt}{Calculated Prism DAPB Energies
as a Function of Temperature.}%
\end{figure}%
%

\begin{figure}[ht]%
%
\centerline{\epsfbox{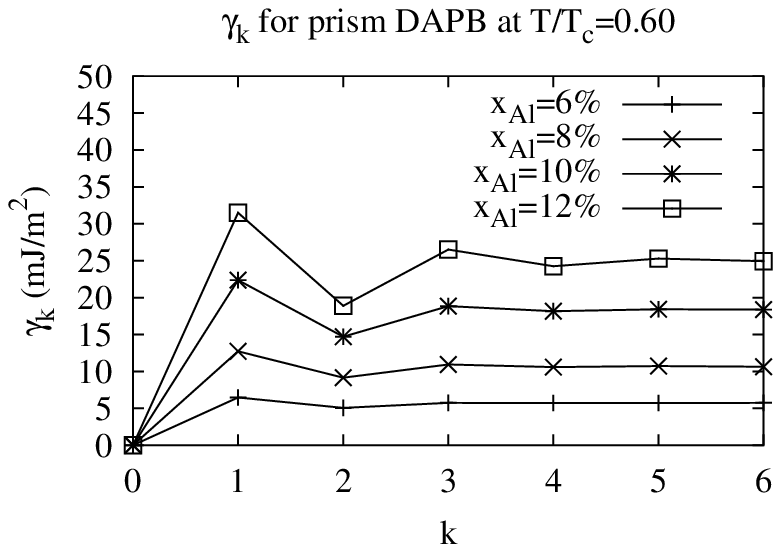}}%
\mynewcaption{figgpx}{Calculated Prism DAPB Energies
as a Function of Concentration.}%
\end{figure}%
%

\begin{figure}[ht]%
%
\centerline{\epsfbox{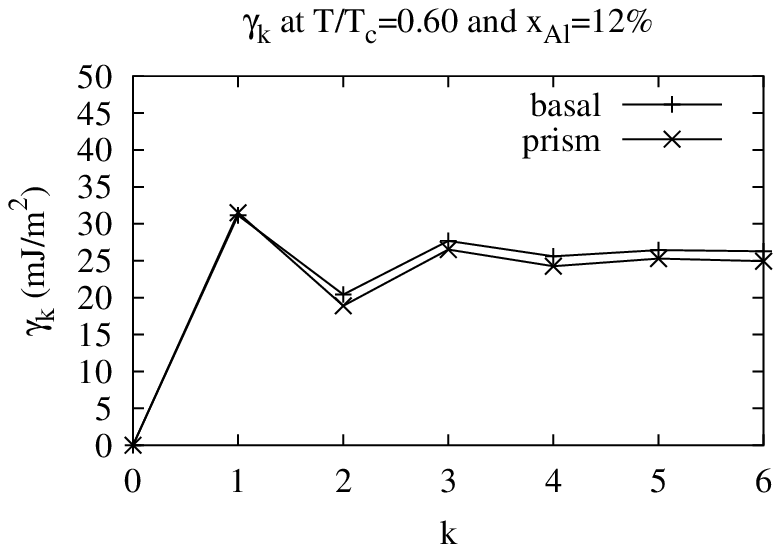}}%
\mynewcaption{figgpb}{Calculated Prism and basal DAPB
Energies.}%
\end{figure}%
%

\clearpage

\renewcommand{\contentsline}[3]{#2}
\renewcommand{\numberline}[2]{\begin{trivlist}\item[Figure #1.] #2\end{trivlist}}
\renewcommand\listfigurename{Figure Captions}

\listoffigures

\end{document}